\documentclass[12pt,dvips,letterpaper]{article}

\usepackage{amsmath}    
\usepackage{amsfonts}   
\usepackage{graphicx}   
\usepackage{verbatim}   
\usepackage{color}      
\usepackage{xcolor}
\usepackage{subfigure}  
\usepackage{hyperref}   
\usepackage{listings}   
\usepackage{epsf}
\usepackage{array}
\usepackage{graphicx}
\usepackage[font={small,it}]{caption}
\usepackage{authblk}

\usepackage{caption}
\DeclareCaptionFont{white}{\color{white}}
\usepackage{courier}    

\usepackage{lipsum} 

\title{Context-Dependent Functions: Narrowing the Realm of Turing's Halting Problem}

\author{Nicholas J. Macias}

\affil{Department of Engineering and Computer Science\\
Clark College\\
Vancouver, WA 98663}

\date {13 Nov 2014}
\DeclareCaptionFormat{listing}{\colorbox{gray}{\parbox{\linewidth}{#1#2#3}}}

\begin{document}
\maketitle

\abstract

This paper describes Turing's Halting Problem (HP), and reviews the classic proof that no function exists that can solve HP. The concept of a ``Context-Dependent Function" (CDF), whose behavior varies based on seemingly irrelevant changes to a program calling that function, is introduced, and the proof of HP's undecidability is re-examined in light of CDFs. The existence of CDFs is established via a pair of examples of such functions. The conclusion of the proof of HP's undecidability is thus shown to be overly strong, as it doesn't show that no solution to HP exists, but rather that a solution must be a CDF. A higher-level analysis of this work is given, followed by conclusions and comments on future work.

\section{Introduction}
In 1928, David Hilbert posed his {\it Entscheidungsproblem}, which asked for an algorithm to determine the universal validity of any first-order logic statement \cite{hilbert}. In studying this question, Alan Turing developed a theoretical model of computation (now called a ``Turing Machine"), and proposed the well-known ``Halting Problem" \cite{HP}. The Halting Problem (HP) is, in effect, the problem of writing a computer program that can determine whether another program will (when given certain input) eventually halt, or will run forever. The problem was shown to be undecidable in 1936 by Alan Turing, using a form of {\em Cantor Diagonalization} \cite{Cantor}.

The Halting Problem (HP) is significant for three reasons: 
\begin{itemize}
\item it was one of the first problems shown to be undecidable;
\item its proof required creating a formal definition of computation, which led to the definition of a Turing Machine; and
\item its undecidability led to conclusions about the decidability of other problems \cite{Sig}.
\end{itemize}

The classic proof that HP is undecidable is based on a contradiction. First, it's assumed that a function $HALT$ solves the halting problem. Next, a second function $BAD$ is written that uses $HALT$ in a particular way. It's then shown that when running $BAD$, $HALT$ fails to correctly predict $BAD's$ own behavior, thus contradicting the statement that $HALT$ solves the Halting Problem. The contradiction means our original assumption was wrong, hence $GOOD$ does not solve the halting problem.

The present work re-examines this proof, calling into question how much can be said about $HALT$ based on the behavior of $BAD$.

\section{A Classic Proof of HP's Undecidability}
A classic proof of HP's undecidability proceeds as follows:
\begin{enumerate}
\item Suppose there is a solution to the halting problem. This means there is a function $HALT(P,I)$ that accepts two arguments:
\begin{itemize}
  \item $P$ is a representation of a program (perhaps as source code or object code); and
  \item $I$ is a set of input to be given to the program represented by $P$;
\end{itemize} and can predict whether or not $P$ eventually halts when run with input $I$. One may assume without loss of generality that $HALT$ returns $1$ if $P$ will halt, and 0 otherwise.

\item Write a new function $BAD$ that uses $HALT$ as follows:

\begin{lstlisting}[label=code2,caption={A Derived Function That Uses $HALT$}]
    BAD(P)  // this is our new function
    begin
      if (HALT(P,P)=0) // HALT says P won't halt
        then halt;   
      while (1=1){} // loop here forever
                    // (HALT said P WILL halt)
    end
\end{lstlisting}

\item Now run $BAD(BAD)$ and consider what happens:
\begin{itemize}
  \item If $HALT(BAD,BAD)$ returns $1$, then $HALT$ has determined that the function ``$BAD$" will halt when run on itself. Since $HALT$ returns $1$ though, $BAD$ then executes the while(1=1) statement, i.e., $BAD$ enters an infinite loop, and never halts. In this case, $HALT$ failed to correctly predict the behavior of $BAD(BAD)$.

  \item If instead $HALT(BAD,BAD)$ returns $0$, then $HALT$ has determined that $BAD$ will not halt. Since $HALT$ returns $0$ though, $BAD$ immediately executes a halt instruction, i.e., $BAD$ halts. In this case also, $HALT$ failed to correctly predict the behavior of $BAD(BAD)$.
\end{itemize}

\item Note that these are the only two possible cases, and in either case (whether $HALT$ returns $0$ or $1$), $HALT's$ behavior is incorrect, i.e., $HALT$ fails to answer the Halting Problem correctly. This contradicts Step 1.

\item Conclude that the assumption made in Step 1 must be incorrect, i.e., ``there is no procedure that solves the halting problem" \cite{Rosen}.
\end{enumerate}

This is the classic method of proving that HP is undecidable. It's a proof that is presented in numerous undergraduate computer science classes, and can be found in many texts on automata theory, foundations of computation, or discrete math. Nonetheless, there is a subtle flaw in this proof. While Step 4 is clearly a contradiction, the proof's subsequent conclusion is overly strong, as explained in the next section.

\section {What the Contradiction Really Means}
The contradiction (reached in Step 4) shows that {\em when used inside the function $BAD$}, $HALT$ does not solve the Halting Problem. Now consider the following function ``$GOOD$":

\begin{lstlisting}[label=code1,caption={A Different Function That Uses $HALT$}]
    GOOD(P,I) // determine program P's behavior
              // when run with input I
    begin
      if (HALT(P,I)=1)
         then print "Program halts."
         else print "Program runs forever."
      halt // program halts after giving answer
    end
\end{lstlisting}

Is it possible that inside {\em this} function, $HALT$ {\em does} solve the Halting Problem, always correctly predicting a function's behavior (including delivering a correct prediction for the behavior of $BAD(BAD)$)?

In general, one expects the answer to be ``no." If $HALT$ worked correctly inside $GOOD$, then it should also work correctly inside $BAD$ (which it doesn't). Simply calling a function and taking action based on its return value shouldn't perturb a function's behavior and somehow cause it to suddenly stop working correctly. And in general, this analysis is correct. But there is a class of computer functions whose behavior is dependent on the context in which they are called or used: these may be called {\em Context-Dependent Functions} (CDFs). With this notion, the above proof does not show that the Halting Problem is undecidable, but shows something slightly weaker: that only a CDF can solve the Halting Problem. If a function $HALT$ is not context-dependent, then the above proof correctly leads to the conclusion that $HALT$ - no matter which function it is used in - cannot be a solution to HP (because it fails in particular to analyze $BAD(BAD)'s$ behavior inside $BAD$). But if $HALT$ is a CDF, it may be able to {\em always} give correct predictions when called from inside $GOOD$, even though it fails to do so when called from inside $BAD$.

Of course, this argument is meaningless unless the class of CDFs is non-empty. The following section gives two examples of context-dependent functions. {\bf These examples have nothing directly to do with a solution to the Halting Problem.} They are merely a proof that the space of CDFs (the only space within which a solution to HP may exist) is non-empty.

\section{Examples of Context-Dependent Behavior of a Function}
Is it possible for a function to work correctly when used certain ways but to fail when used other ways? Of course, if it's called incorrectly, or if something perturbs its code, then its behavior may change; but what about seemingly trivial differences in how a function is used? Is it possible, for example, that simply changing the {\em names of the variables} passed to a function, or adding a print statement {\em after} a function call, can somehow affect the function's behavior, causing it to malfunction? The answer is yes, as the following examples show.

\subsection{First Example of a Context-Dependent Function}

The first CDF presented is $mul(x,y)$, whose purpose is to multiply two integers and return their product (as well as print the values of its arguments). Listing \ref{go1} (``$good.c$") shows a function in which $mul$ behaves correctly; Listing \ref{go2} (``$bad.c$") shows a similar function that uses $mul$ in almost the same way, but in which $mul$ malfunctions.

\begin{lstlisting}[label=go1,caption={good.c}]
#include <stdio.h>
main()
{
  int x,y,z;
  x=12; y=3;
  z=mul(x,y);
  printf("%d*%d=%d\n",x,y,z);
}
\end{lstlisting}

\begin{lstlisting}[label=go2,caption={bad.c}]
#include <stdio.h>
main()
{
  int x,y,z;
  x=3; y=12;
  z=mul(y,x);
  printf("%d*%d=%d\n",y,x,z);
}
\end{lstlisting}

In each case, $mul(12,3)$ is called and the expected return value is $36$. But here are the results of running these functions: 

\begin{lstlisting}[label=gouse,caption={Compilation and Execution of $good$}]
% gcc -o good good.c mul.o
% good
First argument is 12; second argument is 3
12*3=36
%
\end{lstlisting}

\begin{lstlisting}[label=gouse2,caption={Compilation and Execution of $bad$}]
% gcc -o bad bad.c mul.o
% bad
First argument is 12; second argument is 3
12*3=9
%
\end{lstlisting}

Note that this test case is not an isolated example: $mul$ works perfectly inside $good$ no matter what values (barring integer overflow) are passed to it; whereas $mul$ inside $bad$ fails for almost all pairs of integers. This unusual behavior is easily explained by looking at the code for $mul$:

\begin{lstlisting}[label=mul,caption={Function $mul$  Whose Behavior is Context-Dependent}]
#include <stdio.h>
int mul(int xx, int yy)
{
  int *ptr,x,y;
  ptr=&xx;
  ptr+=(0x2c>>2);
  x=*ptr;
  printf("First argument is %d; ",xx);
  printf("second argument is %d\n",yy);
  return(x*yy);
}
\end{lstlisting}

The behavior is not at all mysterious: it uses pointer arithmetic to access the original variable (passed as the first argument) using the {\em local} address of the first argument. Calling $mul(x,y)$ is different from calling $mul(y,x)$ even if the values of $x$ and $y$ are swapped, because of the peculiar way in which $mul$ uses its arguments.

The following is an example of another CDF, with a different type of dependence on its context.

\subsection{Second Example of a Context-Dependent Function}

Below is a second CDF (named ``$mul2$"), whose function is again to multiply two integer arguments and return their product. Consider the following function $good2$ that uses $mul2$:

\begin{lstlisting}[label=good2,caption={good2.c}]
#include <stdio.h>
main()
{
  int x,y,z;
  x=12;
  y=3;
  z=mul2(x,y);
  printf("%d*%d=%d\n",x,y,z);
}
\end{lstlisting}

and a slightly-modified version $bad2$:

\begin{lstlisting}[label=bad2,caption={bad2.c}]
#include <stdio.h>
main()
{
  int x,y,z;
  x=12;
  y=3;
  z=mul2(x,y);
  printf("%d*%d = %d\n",x,y,z);
}
\end{lstlisting}

In both cases, $mul2$ is called {\em identically}, with arguments $(x,y)$, whose values are $(12,3)$. The {\bf only} difference between $good2$ and $bad2$ is in the subsequent print statement (which is executed {\em after} the call to $mul2$): in $bad$, there is a space added before and after the ``=" in the printed string.

Surely such a minor change - made after calling $mul2$ - shouldn't affect the behavior of $mul2$! But it does, as Listings \ref {mul2run1} and \ref{mul2run2} show.

\begin{lstlisting}[label=mul2run1,caption={Compilation and Execution of good2}]
% gcc -o good2 good2.c mul2.o
% good2
12*3=36
%
\end{lstlisting}

\begin{lstlisting}[label=mul2run2,caption={Compilation and Execution of bad2}]
% gcc -o bad2 bad2.c mul2.o
% bad2
12*3 = 7
%
\end{lstlisting}

Listing \ref {mul2} shows the code for $mul2$.

\begin{lstlisting}[label=mul2,caption={Function $mul2$ Whose Behavior is Context-Dependent}]
int mul2(int x, int y)
{
  int z;
  long int *i;
  i=0x100403035;
  z=x*y;
  z=z+(((*i)&0x7f)-'=');
  return(z);
}
\end{lstlisting}

\section{Analysis/Motivation}

One way to view CDFs is to consider the difference between the space of mathematical functions and the space of computer functions. If $f(x)$ is a mathematical function with a certain behavior, say $f(x_0)=y_0$, then $g(f(x_0)) = g(y_0)$ for all functions $g$. But if $f(x)$ is a {\bf computer} function, this may not be the case: it may be that for some function $g$, $g(f(x_0)) \neq g(y_0)$, even though $f(x_0)=y_0$.

As for the meaning of this present work, it is only an existential demonstration of the {\em possibility} of a solution to the Halting Problem. Since a CDF may work correctly inside one function while failing to work correctly inside a different function, the conclusion of the proof of HP's undecidability is only valid if one assumes that $HALT$ is not a CDF. If instead $HALT$ is context-dependent, then while it clearly fails to solve HP when called inside $BAD$, it's possible that it always succeeds in solving HP when called inside $GOOD$. Thus, the proof given in Section 2 does not prove that HP is undecidable: it only shows that a function that solves the Halting Problem must be a context-dependent function. As the prior section shows that CDFs exist, this raises the possibility that the Halting Problem can be solved.

It should be noted that these examples of CDFs - particularly $mul2$ - did not arise from an attempt to find fault with the proof of HP's undecidability. Rather, they came from thought experiments on what the nature of a solution to HP might be {\em if it existed} (while not actually believing the existence of such a solution). In considering this question, it occurs that a potential solution will likely need to be able to analyze itself, e.g. by setting pointers to its own code or data space and resolving them. The expected vulnerability of such code to seemingly minor changes in a calling function is what led to the notion of a CDF and its role in analyzing the proof of HP's undecidability.

To see how a solution to HP might make use of such analysis, consider the following high-level pseudocode for a hypothetical $HALT(P,I)$ function to solve HP. $HALT$ would need to do the following:
\begin{enumerate}
\item analyze the code from which it was called, and if the calling context is anything other than the exact code for $GOOD$, then execute a halt;
\item analyze the program $P$ with input $I$ (how to do this is of course a huge unknown!), and if $P$ ever calls $HALT$, then return 1;
\item analyze $P$ with input $I$, and if it ever executes a halt instruction, then return 1;
\item else return 0.
\end{enumerate}

Step 1 handles the case where $HALT$ is used inside any program other than $GOOD$. In that case, the program will halt (note that this is not referring to the program that $HALT$ is analyzing: it's referring to the program that caused $HALT$ to be executed).

Step 2 is only reached if $HALT$ is being run from inside $GOOD$, i.e., from a context in which it is expected to correctly analyze $(P,I)$. In Step 2, if $P$ calls $HALT$ then there are only two possibilities. Either:
	\begin{itemize}
	\item $P$ is the program $GOOD$, which by design will eventually halt, in which case $HALT$ returns the correct value ($1$); or
	\item $P$ is some other program that calls $HALT$, which by Step 1 will halt (when $HALT$ halts inside it), in which case again $HALT$ returns the correct value ($1$).
	\end{itemize}

Steps 3 and 4 run in all other cases, i.e., cases where $GOOD$ is analyzing a program $P$ which does not call $HALT$.

The above algorithm thus handles the paradox-inducing cases, leaving $HALT$ with the simpler (though phenomenally-complex and likely impossible) task of analyzing a program that does not call $HALT$.

Observe that $HALT$ is a context-dependent function: its behavior changes depending on the context from which it is called. Also, note that in some calling contexts, $HALT$ is {\em not} a solution to the halting problem: when called from inside $BAD$ for example, $HALT$ does not return a 0 or a 1, but instead halts itself. Moreover, if $GOOD$ is called from another program, then $HALT$ will fail to return a 0 or a 1. In each of these cases, $HALT$ does not solve HP; {\em it's the combination of $GOOD$ and $HALT$ that solves the halting problem.}

While the details of ``analyzing program $P$ with input $I$" are unknown (this is, after all, the crux of the Halting Problem), the above algorithm at least handles the self-referential cases that deliberately trip up the algorithm in the classic proof of HP's undecidability.

\section{Conclusions and Future Work}
The primary result of this work is that the standard conclusion of the proof given in Section 2 - that an HP solver $HALT$ cannot exist - is too strong of a conclusion. Instead, the weaker conclusion - that $HALT$ must be a context-dependent function - should be accepted, thus allowing the possibility that $HALT$ {\em does} in fact work correctly, inside a carefully chosen calling function $GOOD$.

While the present work makes no claims to solve the Halting Problem -- and gets one no closer to an actual solution -- it does at least re-open the {\em theoretical} possibility of a solution. Moreover, given the similarity between work on the undecidability of the Halting Problem and the proof of G\"{o}del's incompleteness theorem \cite{Incomplete}, it may be interesting to re-examine the latter in light of the present work.

\bibliography{paper}
\bibliographystyle{plain}

\end{document}